\documentstyle[11pt,aaspp4]{article}

\newcommand{\wisk}[1]{\ifmmode{#1}\else{$#1$}\fi}

\def\gsim{\;_\sim^>\;}
\def\lsim{\;_\sim^<\;}

\begin{document}

\vspace{-2.0truecm}
\begin{flushright}
KSUPT-99/1, KUNS-1552 \hspace{0.5truecm} February 1999
\end{flushright}
\vspace{-0.5truecm}

\title{Python I, II, and III CMB Anisotropy Measurement Constraints on
  Open and Flat-$\Lambda$ CDM Cosmogonies}

\author{
  Gra{\c c}a~Rocha\altaffilmark{1,2},
  Rados{\l}aw~Stompor\altaffilmark{3,4},
  Ken~Ganga\altaffilmark{5},  
  Bharat~Ratra\altaffilmark{1}, 
  Stephen~R.~Platt\altaffilmark{6},  
  Naoshi~Sugiyama\altaffilmark{7},
  and
  Krzysztof~M.~G\'orski\altaffilmark{8,9}
  }

\altaffiltext{1}{Department of Physics, Kansas State University,
                 Manhattan, KS 66506.}
\altaffiltext{2}{Centro de Astrof\'{\i}sica da Universidade do Porto, 
                 Rua das Estrelas s/n, 4100 Porto, Portugal.}
\altaffiltext{3}{Center for Particle Astrophysics, University of 
                 California, Berkeley, CA 94720.}
\altaffiltext{4}{Copernicus Astronomical Center, Bartycka 18, 
                 00-716 Warszawa, Poland.}
\altaffiltext{5}{Infrared Processing and Analysis Center, MS 100--22, 
                 California Institute of Technology, Pasadena, CA 91125.}
\altaffiltext{6}{Snow and Ice Research Group, University of Nebraska,
                 Lincoln, NE 68583-0850.}
\altaffiltext{7}{Department of Physics, Kyoto University,
                 Kitashirakawa-Oiwakecho, Sakyo-ku, Kyoto 606-8502, Japan.}
\altaffiltext{8}{Theoretical Astrophysics Center, Juliane Maries Vej 30,
                 2100 Copenhagen \O, Denmark.}
\altaffiltext{9}{Warsaw University Observatory, Aleje Ujazdowskie 4, 
                 00-478 Warszawa, Poland.}

\begin{abstract}
  We use Python I, II, and III cosmic microwave background anisotropy
  data to constrain cosmogonies. We account for the Python beamwidth
  and calibration uncertainties.
  We consider open and spatially-flat-$\Lambda$ cold dark matter cosmogonies,
  with nonrelativistic-mass density parameter $\Omega_0$ in the 
  range 0.1--1, baryonic-mass density parameter $\Omega_B$ in the range
  (0.005--0.029)$h^{-2}$, and age of the universe $t_0$ in the range   
  (10--20) Gyr. Marginalizing over all parameters but $\Omega_0$, the 
  combined Python data favors an open (spatially-flat-$\Lambda$) model with 
  $\Omega_0\simeq$ 0.2 (0.1).
  At the 2 $\sigma$ confidence level model normalizations deduced from the
  combined Python data are mostly consistent with those drawn from the 
  DMR, UCSB South Pole 1994, ARGO, MAX 4 and 5, White Dish, and SuZIE data 
  sets. 
\end{abstract}

\keywords{cosmic microwave background---cosmology: observations---large-scale
  structure of the universe}

\section{Introduction}

Ganga et al. (1997a, hereafter GRGS) developed a technique to account for 
uncertainties, such as those in the beamwidth and the calibration, in 
likelihood analyses of cosmic microwave background (CMB) anisotropy data. 
This technique has been used in conjunction with theoretically-predicted CMB 
anisotropy spectra in analyses of the 
Gundersen et al. (1995) UCSB South Pole 1994 data, the Church et al. (1997)
SuZIE data, the MAX 4+5 data (Tanaka et al. 1996; Lim et al. 1996), the 
Tucker et al. (1993) White Dish data, and the de Bernardis et al. (1994)
ARGO data (GRGS; Ganga et al. 1997b, 1998; Ratra et al. 1998, 1999a, hereafter
R99a). A combined analysis of all these data sets is presented in Ratra et al.
(1999b, hereafter R99b).

In this paper we present a similar analysis of CMB anisotropy data from the 
Python I, II, and III observations performed at the South Pole (Dragovan et 
al. 1994, hereafter D94; Ruhl et al. 1995b, hereafter R95; Platt et al. 1997,
hereafter P97). The Python detectors and telescope are described by Ruhl (1993) 
and D94; also see Ruhl et al. (1995a) and Alvarez (1996). In what follows we
review the information needed for our analysis.

Python I, II, and III CMB data were taken in a frequency band centered at 
90 GHz with four bolometric detectors centered at the corners of a 
$2.\!\!^\circ 75$ by $2.\!\!^\circ 75$ square on the sky. The beam profiles 
are well-approximated by a Gaussian of FWHM 
$0.\!\!^\circ 75 \pm 0.\!\!^\circ 05$ (one standard deviation uncertainty).
Observations were centered at $\alpha = 23.\!\!^{\rm h}37$, 
$\delta = -49.\!\!^\circ 44$ (J2000.0). Python I and II data were taken at a 
single telescope elevation. Python III data were taken at this fiducial 
elevation as well as two additional elevations offset $2.\!\!^\circ 75/3$ 
on the sky above 
and below the fiducial elevation. The reduced Python data are shown in Figure 1.

All of the Python measurements were made by switching the four beams
horizontally across the sky in a three-point pattern by rotating a vertical
flat mirror at 2.5 Hz.  This chopping pattern was then combined with slow
(typically 0.1 Hz) azimuthal beam switching of the entire telescope to
produce a four-beam response to a sky signal.

The chopper throw and azimuthal telescope beam switching were both 
$2.\!\!^\circ 75$ on the sky for the Python I and II observations.  Python I 
(hereafter I) resulted in 16 data points, 8 each from the lower and 
upper rows of detectors (D94).  Python II observed two sets of points on the 
sky (R95), although one of the detectors did not work during this season.  
The first set of observations, hereafter IIA, overlapped the I points and 
yielded 7 measurements from the lower row of detectors and 8 from the upper 
row.  The second set, hereafter IIB, measured the same number of points on the sky as the first set, but these were offset in azimuth relative to the I 
points by $-2.\!\!^\circ 75/2$ on the sky.

Two series of measurements were made at each of the three elevations
observed during the Python III season (P97). For each series, the physical throw
of the chopper was the same at all elevations.  Because the actual throw on
the sky depends on elevation, the Python III beam was smeared to a Gaussian
FWHM of $0.\!\!^\circ 82 \pm 0.\!\!^\circ 05$ (one standard deviation 
uncertainty) when forming the four-beam pattern to account for the imperfect 
overlap of the beams caused by this effect. This procedure also accounted for 
the relative pointing uncertainty and fluctuations in the chopper throw.

The first series of Python III measurements, hereafter IIIL,
used the same chopper and beam switch parameters as Python I and II.  The
IIIL data consists of two sets of points taken at the I elevation but
offset in right ascension by  $-2.\!\!^\circ 75/3$ and 
$-2 \times 2.\!\!^\circ 75/3$ on the sky. Two sets of measurements were also 
made at each of the lower and upper elevations, but these were offset in right 
ascension by $0.\!\!^\circ 0$ and $+ 2 \times 2.\!\!^\circ 75/3$ on the sky 
relative to I.

The second series of Python III measurements, hereafter IIIS, were made
with both the chopper throw and telescope beam switch reduced to 
$2.\!\!^\circ 75/3$. For each of the three telescope elevations, the IIIS data
consists of points separated horizontally by $2.\!\!^\circ 75/3$ on
the sky for each of the 2 rows of detectors. Figure 1 shows the points
observed by IIIL and IIIS at each elevation. Together, I, II, and III
densely sample a $5.\!\!^\circ 5$ by $22^\circ$ region of the sky.

The 1 $\sigma$ absolute calibration uncertainty in the Python data is $20\%$
(D94; R95; P97) and is accounted for in our analysis. The absolute pointing
uncertainty is $0.\!\!^\circ 1$ (D94; R95) and is not accounted for in our 
analysis.

In $\S$2 we summarize the computational techniques used in our analysis. See 
GRGS and R99a for detailed discussions. Results are presented and discussed 
in $\S$3. Conclusions are given in $\S$4.

\section{Summary of Computation}

The zero-lag window function for the Python observations are shown in Figure 2 
and the zero-lag window function parameters are in Table 1.

In this paper we focus on a spatially open CDM model and
a spatially flat CDM model with a cosmological constant $\Lambda$. These
low density models are largely consistent with current observational 
constraints.\footnote{
While not considered in this paper, a time-variable cosmological ``constant" 
dominated spatially-flat model is also largely consistent with
current data (e.g., Peebles \& Ratra 1988; Sugiyama \& Sato 1992; Ratra \& 
Quillen 1992; Frieman \& Waga 1998; Ferreira \& Joyce 1998; Wang \& Steinhardt 
1998; Carroll 1998; Hu et al. 1999; Huterer \& Turner 1999; Liddle \& 
Scherrer 1999; Starobinsky 1998).}
For recent discussions see Park et al. (1998), Retzlaff et al. 
(1998), Croft et al. (1999), and Peebles (1999).

The models have Gaussian, adiabatic primordial energy-density power
spectra. The flat-$\Lambda$ model CMB anisotropy computations use a 
scale-invariant energy-density perturbation power spectrum (Harrison 1970;
Peebles \& Yu 1970; Zel'dovich 1972), as predicted in the simplest 
spatially-flat inflation models (Guth 1981; Kazanas 1980; Sato 1981).
The open model computations use the energy-density power spectrum 
(Ratra \& Peebles 1994, 1995; Bucher, Goldhaber, \& Turok 1995; Yamamoto,
Sasaki, \& Tanaka 1995) predicted in the simplest open-bubble inflation 
models (Gott 1982). The computation of the CMB anisotropy spectra is described  by Stompor (1994) and Sugiyama (1995). 

As discussed in R99a, the spectra are parameterized by their quadrupole-moment
amplitude $Q_{\rm rms-PS}$, the nonrelativistic-mass density parameter 
$\Omega_0$, the baryonic-mass density parameter $\Omega_B$, and the age of the 
universe $t_0$. The spectra are computed for a range of $\Omega_0$ spanning the 
interval 0.1 to 1 in steps of 0.1, for a range of $\Omega_B h^2$ [the Hubble 
parameter $h = H_0/(100\ {\rm km}\ {\rm s}^{-1}\ {\rm Mpc}^{-1})$] spanning 
the interval 0.005 to 0.029 in steps of 0.004, and for a range of $t_0$ 
spanning the interval 10 to 20 Gyr in steps of 2 Gyr. In total 798 spectra were 
computed to cover the cosmological-parameter spaces of the open and 
flat-$\Lambda$ models. Figure 2 shows examples
of the CMB anisotropy spectra used in our analysis. Other examples are in
Figure 2 of R99a and Figure 1 of R99b. 

While it is of interest to also consider other cosmological parameters,
such as tilt or gravity wave fraction or a time-variable cosmological 
``constant" (instead of a constant $\Lambda$), to make the problem tractable we
have focussed on the four parameters mentioned above. We emphasize however that 
the results of the analysis are model dependent. For instance, a time-variable
$\Lambda$ model would likely lead to a different constraint on $\Omega_0$ than 
that derived below in the constant $\Lambda$ model.

Following GRGS, for each of the 798 spectra considered the ``bare" likelihood 
function is computed at the nominal beamwidth and calibration, as well as at 
a number of other values of the beamwidth and calibration determined from the 
measurement uncertainties. The likelihood function used in the derivation of 
the central values and limits is determined by integrating (marginalizing) the 
bare likelihood function over the beamwidth and calibration uncertainties with
weights determined by the measured probability distribution functions of the
beamwidth and the calibration. See GRGS for a more detailed discussion. When 
marginalizing over the beamwidth uncertainty we have checked in a few 
selected cases that the five-point Gauss-Hermite quadrature summation 
approximation to the integral agrees extremely well with the three-point
Gauss-Hermite approximation used by GRGS (and for most of the analysis
in this paper). The likelihoods are a function of four parameters mentioned
above: $Q_{\rm rms-PS}$, $\Omega_0$, $\Omega_B h^2$, and $t_0$. We also compute 
marginalized likelihood functions by integrating over one or more of these 
parameters after assuming a uniform prior in the relevant parameters.
The prior is set to zero outside the ranges considered for the parameters.
GRGS and R99a describe the prescription used to determine central values and 
limits from the likelihood functions. In what follows we consider 1, 2, and
3 $\sigma$ highest posterior density limits which include 68.3, 95.4, and
99.7\% of the area.

\section{Results and Discussion}

Table 2 lists the derived values of $Q_{\rm rms-PS}$ and bandtemperature 
$\delta T_l$
for the flat bandpower spectrum, for various combinations of the I, II, and III
data. These numerical values account for the beamwidth and calibration 
uncertainties. The last two $\delta T_l$ entries in Table 2 are quite
consistent with those derived by P97; the small differences reflect the
different methods used to account for beamwidth and calibration uncertainties
here and in P97.

For the flat bandpower spectrum the combined I, II, and III data average 
1 $\sigma$ $\delta T_l$ error bar is $\sim 25\%$\footnote{
For comparison, the corresponding DMR 1 $\sigma$ $\delta T_l$ error bar is 
$\sim 10-12\%$ (depending on model, G\'orski et al. 1998).} 
: Python data results in a very 
significant detection of CMB anisotropy, even after accounting for beamwidth 
and calibration uncertainties. Note that the calibration uncertainty, $20\%$, 
is the most important contributor to this error bar. 

A number of other interesting conclusions follow from the entries in Table 2.
Comparing the I+II and IIIL results, which are from experiments which probe 
almost identical angular scales, we see that the IIIL amplitude is 
$\sim 1$ $\sigma$ higher than the I+II amplitude. Comparing the result from
the analysis of the coadded I and IIA data (which are from experiments with
identical window functions) and the result from the full analysis of the 
I and IIA data (which takes into account all the spatial correlations), 
we see that the deduced amplitudes are almost identical. This is probably 
mostly a reflection of the fact that the individual I and IIA amplitudes 
are almost identical (see Table 2).

As discussed in R99a and R99b, the four-dimensional posterior probability 
density distribution function $L(Q_{\rm rms-PS}, \Omega_0, \Omega_B h^2, t_0)$ 
is nicely peaked in the $Q_{\rm rms-PS}$ direction but fairly flat in the other 
three directions. Marginalizing over $Q_{\rm rms-PS}$ results in a 
three-dimensional posterior distribution $L(\Omega_0,  \Omega_B h^2, t_0)$ 
which is steeper, but still relatively flat. As a consequence, limits 
derived from the four- and three-dimensional posterior distributions are
generally not highly statistically significant. We therefore do not show 
contour plots of these functions here. Marginalizing over $Q_{\rm rms-PS}$ and 
one other parameter results in two-dimensional posterior probability 
distributions which are more peaked. See Figures 3 and 4. As in the ARGO (R99a) 
and combination (R99b) data set analyses, in some cases these
peaks are at an edge of the parameter range considered.

Figure 3 shows that the two-dimensional posterior distributions allow one to 
distinguish between different regions of parameter space at a fairly high 
formal level of confidence. For instance, the open model near $\Omega_0 \sim 0.75$, $\Omega_B h^2 \sim 0.03$, and $t_0 \sim 20$ Gyr, and the flat-$\Lambda$ 
model near $\Omega_0 \sim 0.6$, $\Omega_B h^2 \sim 0.03$, and $t_0 \sim 20$ Gyr,
are both formally ruled out at $\sim 3$ $\sigma$ confidence. However, we 
emphasize, as discussed in R99a and R99b, care must be exercised when 
interpreting the discriminative power of these formal limits, 
since they depend sensitively on the fact that the uniform prior has been set 
to zero outside the range of the parameter space we have considered.

Figure 4 shows the contours of the two-dimensional posterior distribution 
for $Q_{\rm rms-PS}$ and $\Omega_0$, derived by marginalizing the 
four-dimensional distribution over $\Omega_B h^2$ and $t_0$. These are shown
for the combined Python I, II, and III data, the DMR data, and three 
combinations of data from the SP94, ARGO, MAX 4+5, White Dish, and SuZIE 
experiments (R99b), for both the open and flat-$\Lambda$ models. Constraints
on these parameters from the combined Python data are consistent with those 
from the DMR data for the flat-$\Lambda$ models, panel $a)$, while for the 
open model, panel $b)$, consistency at 2 $\sigma$ (1 $\sigma$) requires
$\Omega_0 \gsim$ 0.2 (0.35). The combined Python data amplitudes are a 
little higher than those derived from the other small-scale data combinations,
panels $c)-h)$, but at 2 $\sigma$ confidence the various amplitudes are
mostly consistent.

Figure 5 shows the one-dimensional posterior distribution functions for 
$\Omega_0$, $\Omega_B h^2$, $t_0$, and $Q_{\rm rms-PS}$, derived by 
marginalizing the four-dimensional posterior distribution over the other three 
parameters. From these one-dimensional distributions, the combined I, II, and
III data favors an open (flat-$\Lambda$) model with $\Omega_0$ = 0.19 (0.10), 
or $\Omega_B h^2$ = 0.005 (0.005), or $t_0$ = 10 (13) Gyr, amongst the models
considered. At 2 $\sigma$ confidence the combined Python data formally rule out 
only small regions of parameter space. From the one-dimensional distributions
of Figure 5, the data requires $\Omega_0$ $< 0.71$ or $> 0.8$
($\Omega_0$ $< 0.55$ or $> 0.63$), or $\Omega_B h^2$ $< 0.028$ 
($\Omega_B h^2$ $< 0.028$), or $t_0$ $< 20$ Gyr ($t_0$ $< 20$ Gyr) for the open (flat-$\Lambda$) model at 2 $\sigma$. As discussed in R99a and R99b, 
care is needed when interpreting the discriminative power of these formal
limits. These papers also discuss a more conservative Gaussian posterior
distribution limit prescription. Using this more conservative prescription, we 
find only an upper 1 $\sigma$ limit on $\Omega_0\ (\lsim\ 0.5)$ in the open 
model. 

While the statistical significance of the constraints on cosmological 
parameters is not high, it is reassuring that the combined Python data 
favor low-density, low $\Omega_B h^2$, young models, consistent with some
of the indications from the combinations of CMB anisotropy data considered 
by R99b, and the indications from most recent non-CMB observations (see
discussion in R99b).

The peak values of the one-dimensional posterior distributions shown in 
Figure 5 are listed in the figure caption for the case when the 
four-dimensional posterior distributions are normalized such that
$L(Q_{\rm rms-PS}\ =\ 0\ \mu{\rm K})\ =\ 1$. With this normalization, 
marginalizing over the remaining parameter the fully marginalized
posterior distributions are $2\times 10^{106}(1\times 10^{106})$ for the 
open (flat-$\Lambda$) model and the combined Python data. This is qualitatively
consistent with the indication from panels $a)$ and $b)$ of Figure 5 that the 
most-favored open model is somewhat more favored than the most-favored
flat-$\Lambda$ one.

\section{Conclusion}

The combined Python I, II, and III data results derived here are mostly 
consistent with those derived from the DMR, SP94, ARGO, MAX 4+5, White Dish and 
SuZIE data. The combined Python data significantly constrains $Q_{\rm 
rms-PS}$ (for the flat bandpower spectrum $Q_{\rm rms-PS}\ = \ 40{+12 \atop
-8}\ \mu$K at 1 $\sigma$) and weakly favors low-density, low $\Omega_B h^2$, young models.

\bigskip

We acknowledge helpful discussions with D. Alvarez, M. Dragovan, G. Griffin, 
J. Kovac, and J. Ruhl. This work was partially carried out at the Infrared 
Processing and Analysis Center and the Jet Propulsion Laboratory, 
California Institute of Technology, under a contract with the National 
Aeronautics and Space Administration. KG also acknowledges support from 
NASA ADP grant NASA-1260. BR and GR acknowledge support from NSF grant 
EPS-9550487 with matching support from the state of Kansas and from a 
K$^*$STAR First award. GR also acknowledges support from a PRAXIS XXI program
of FCT (Portugal) grant. RS acknowledges support from NASA AISRP grant 
NAG-3941 and help from Polish Scientific Committee (KBN) grant 2P03D00813.

\clearpage

\begin{table}
\begin{center}
\caption{Numerical Values for the Zero-Lag Window Function 
Parameters\tablenotemark{a}}
\vspace{0.3truecm}
\tablenotetext{{\rm a}}{The value of $l$ where $W_l$ is
largest, $l_{\rm m}$, the two values of $l$ where $W_{l_{e^{-0.5}}} =
e^{-0.5} W_{l_{\rm m}}$, $l_{e^{-0.5}}$, the effective multipole,
$l_{\rm e} = I(lW_l)/I(W_l)$, and 
$I(W_l) = \sum^\infty_{l=2}(l+0.5)W_l/\{l(l+1)\}$.}
\begin{tabular}{lccccc}
\tableline\tableline
  & $l_{e^{-0.5}}$ & $l_{\rm e}$ & $l_{\rm m}$
  & $l_{e^{-0.5}}$ & $\sqrt{I(W_l)}$  \\
\tableline
Python I/II &  53 &  91.7 &  73 &  99 & 1.34  \\
Python IIIL &  52 &  87.7 &  72 &  98 & 1.30  \\
Python IIIS & 128 & 171   & 176 & 230 & 0.623 \\
\tableline
\end{tabular}
\end{center}
\end{table}

\clearpage

\begin{table}
\begin{center}
\caption{Numerical Values for $Q_{\rm rms-PS}$ and $\delta T_l$ from Likelihood
Analyses Assuming a Flat Bandpower Spectrum}
\vspace{0.3truecm}
\tablenotetext{{\rm a}}{C(...+...) means that the data from the two sets of observations
have been coadded prior to analysis, II refers to the combined IIA and IIB 
data, and III refers to the combined IIIL and IIIS data.} 
\tablenotetext{{\rm b}}{For each data set, the first of the three entries is 
where the posterior probability density distribution function peaks and the 
vertical pair of numbers are the $\pm 1$ $\sigma$ (68.3\% highest posterior
density) values.} 
\tablenotetext{{\rm c}}{Average absolute error on $Q_{\rm rms-PS}$ in 
$\mu$K.}
\tablenotetext{{\rm d}}{Average fractional error, as a fraction of the 
central value.}
\tablenotetext{{\rm e}}{Likelihood ratio.}
\tablenotetext{{\rm f}}{IIB does not have a 2 $\sigma$ highest posterior
density detection; the appropriate equal tail 2 $\sigma$ upper limits
are 50 $\mu$K ($Q_{\rm rms-PS}$) and 77 $\mu$K ($\delta T_l$).}
\begin{tabular}{lccccc}
\tableline\tableline
Data Set\tablenotemark{a} & $Q_{\rm rms-PS}$\tablenotemark{b} &
Ave. Abs. Err.\tablenotemark{c} & Ave. Frac. Err.\tablenotemark{d} &
$\delta T_l$\tablenotemark{b} & LR\tablenotemark{e} \\
{\ }  & ($\mu$K) & ($\mu$K) & {\ } & ($\mu$K) & {\ } \\
\tableline
\medskip
I                    & 39 ${56 \atop 28}$ & 14 & 36\% & 61 ${87 \atop 43}$ & $2 \times 10^{14}$ \\
\medskip
IIA                  & 39 ${57 \atop 27}$ & 15 & 38\% & 60 ${88 \atop 43}$ & $9 \times 10^{12}$ \\
\medskip
C(I+IIA)             & 40 ${57 \atop 29}$ & 14 & 35\% & 62 ${89 \atop 45}$ & $3 \times 10^{31}$ \\
\medskip
I+IIA                & 40 ${56 \atop 29}$ & 14 & 35\% & 61 ${87 \atop 45}$ & $2 \times 10^{31}$ \\
\medskip
IIB\tablenotemark{f} & 9.3 ${20 \atop 0}$ & 10 & 110\% & 14 ${31 \atop 0}$ & $1
$ \\
\medskip
II                   & 31 ${44 \atop 23}$ & 10 & 34\% & 48 ${68 \atop 35}$ & $1 \times 10^{12}$ \\
\medskip
C(I+IIA)+IIB         & 34 ${47 \atop 26}$ & 11 & 31\% & 53 ${73 \atop 40}$ & $1 \times 10^{30}$ \\
\medskip
I+II                 & 34 ${46 \atop 25}$ & 11 & 31\% & 52 ${72 \atop 39}$ & $1 \times 10^{30}$ \\
\medskip
IIIL                 & 41 ${55 \atop 32}$ & 11 & 28\% & 64 ${84 \atop 49}$ & $3 \times 10^{33}$ \\
\medskip
IIIS                 & 42 ${56 \atop 33}$ & 11 & 27\% & 66 ${87 \atop 51}$ & $2 \times 10^{29}$ \\
\medskip
III                  & 41 ${54 \atop 33}$ & 10 & 25\% & 64 ${83 \atop 51}$ & $6 \times 10^{68}$ \\
\medskip
I+II+IIIL            & 39 ${51 \atop 31}$ & 10 & 27\% & 60 ${80 \atop 47}$ & $3 \times 10^{67}$ \\
\medskip
I+II+III             & 40 ${52 \atop 32}$ & 10 & 25\% & 63 ${81 \atop 50}$ & $1 \times 10^{105}$ \\
\tableline
\end{tabular}
\end{center}
\end{table}


\clearpage
\centerline{\bf Figure Captions}

\medskip
\noindent
Fig.~1.--
{\protect Measured thermodynamic temperature differences (with $\pm 
    1$-$\sigma$ error bars). Panel $a)$ shows Python I (open circles), IIA 
    (filled
    squares, offset horizontally from true positions for clarity), IIB 
    (open squares), and IIIL (crosses) data, while panel $b)$ 
    shows Python IIIS data (crosses).}

\medskip
\noindent
Fig.~2.--
{\protect CMB anisotropy multipole moments $l(l+1)C_l/(2\pi )\times
    10^{10}$ (solid lines, scale on left axis, note that these are fractional 
    anisotropy moments and thus dimensionless) as a function of
    multipole $l$, for selected models normalized to the DMR maps
    (G\'orski et al. 1998; Stompor 1997). Panels $a)-c)$ show selected
    flat-$\Lambda$ models. The heavy lines are the $\Omega_0 = 0.1$,
    $\Omega_B h^2 = 0.005$, and $t_0 = 12$ Gyr case, which is close to where 
    the combined Python data likelihoods (marginalized over all but one 
    parameter at
    a time) are at a maximum. Panel $a)$ shows five $\Omega_B h^2$ = 0.005,
    $t_0$ = 12 Gyr models with $\Omega_0$ = 0.1, 0.3, 0.5, 0.7, and 0.9
    in descending order at the $l \sim 200$ peaks. Panel $b)$ shows seven 
    $\Omega_0$ = 0.1, $t_0$ = 12 Gyr models with $\Omega_B h^2$ = 0.029,
    0.025, 0.021, 0.017, 0.013, 0.009, and 0.005 in descending order at
    the $l \sim 200$ peaks. Panel $c)$ shows six $\Omega_0$ = 0.1, 
    $\Omega_B h^2$ = 0.005 models with $t_0$ = 20, 18, 16, 14, 12, and
    10 Gyr in descending order at the $l \sim 200$ peaks. Panels $d)-f)$ 
    show selected open models. The heavy lines are the $\Omega_0 = 0.2$,
    $\Omega_B h^2 = 0.005$, and $t_0 = 10$ Gyr case, which is close to where 
    the combined Python data likelihoods (marginalized over all but one 
    parameter at
    a time) are at a maximum. Panel $d)$ shows five $\Omega_B h^2$ = 0.005,
    $t_0$ = 10 Gyr models with $\Omega_0$ = 1, 0.8, 0.6, 0.4, and 0.2
    from left to right at the peaks (the peak of the $\Omega_0$ = 0.2
    model is off scale). Panel $e)$ shows seven 
    $\Omega_0$ = 0.2, $t_0$ = 10 Gyr models with $\Omega_B h^2$ = 0.029,
    0.025, 0.021, 0.017, 0.013, 0.009, and 0.005 in descending order at
    $l \sim 400$. Panel $f)$ shows six $\Omega_0$ = 0.2, 
    $\Omega_B h^2$ = 0.005 models with $t_0$ = 20, 18, 16, 14, 12, and
    10 Gyr in descending order at $l \sim 400$.
    Also shown are the Python zero-lag window functions $W_l$ (scale on right 
    axis): I/II (long-dashed lines), IIIL (short-dashed lines), and IIIS 
    (dotted lines). See Table 1 for $W_l$-parameter values.}

\medskip
\noindent
Fig.~3.--
{\protect Confidence contours and maxima of the combined Python data 
   two-dimensional posterior 
   probability density distribution functions, as a function of the two 
   parameters on the axes of each panel (derived by marginalizing the 
   four-dimensional posterior distributions over the other two parameters). 
   Dashed lines (crosses) show the contours (maxima) of the open case and 
   solid lines (solid circles) show those of the flat-$\Lambda$ model. Panel 
   $a)$ shows the $(\Omega_B h^2,\ \Omega_0)$ plane, and panel $b)$ shows the 
   $(t_0, \ \Omega_0)$ plane.}

\medskip
\noindent
Fig.~4.-- 
{\protect Confidence contours and maxima of the two-dimensional 
   $(Q_{\rm rms-PS}, \Omega_0)$ posterior probability density distribution 
   functions. Panels $a)$, $c)$, $e)$, \&\ $g)$ in the left column show the 
   flat-$\Lambda$ model and panels $b)$, $d)$, $f)$, \&\ $h)$ in the right 
   column show the open model. Note the different scale on the vertical 
   $(Q_{\rm rms-PS})$ axes of pairs of panels in each row. Heavy lines show
   the $\pm1$ and $\pm2$ $\sigma$ confidence limits and solid circles show the 
   maxima of the two-dimensional posterior distributions derived from the 
   combined Python I, II, and III data. Shaded regions show the two-dimensional 
   posterior distribution 1 $\sigma$ (denser shading) and 2 $\sigma$ (less
   dense shading) confidence regions for the 
   DMR data (G\'orski et al. 1998; Stompor 1997) in panels $a)$ \&\ $b)$;
   for the SP94, ARGO, MAX 4 and 5, White Dish and SuZIE data combination
   (R99b) in panels $c)$ \&\ $d)$; for the previous data combination
   excluding SuZIE (R99b) in panels $e)$ \&\ $f)$; and for the SP94Ka, 
   MAX 4 ID, and MAX 5 HR data combination (R99b) in panels $g)$ \&\ $h)$. 
   The DMR results are a composite of 
   those from analyses of the two extreme data sets: i) galactic frame with 
   quadrupole included and correcting for faint high-latitude galactic 
   emission; and ii) ecliptic frame with quadrupole excluded and no other 
   galactic emission correction (G\'orski et al. 1998). In panels $c)-h)$
   crosses show the maxima of the appropriate non-Python data 
   two-dimensional posterior distributions.}

\medskip
\noindent
Fig.~5.--
{\protect One-dimensional posterior probability density distribution 
   functions for $\Omega_0$, $\Omega_B h^2$, $t_0$, and $Q_{\rm rms-PS}$ 
   (derived by marginalizing the four-dimensional one over the other
   three parameters) in the open and flat-$\Lambda$ models.  These have
   been renormalized to unity at the peaks. Dotted vertical lines show the 
   confidence limits derived from these one-dimensional posterior 
   distributions and solid vertical lines in panels $g)$ and $h)$ show the 
   $\pm 1$ and $\pm 2$ $\sigma$ confidence limits derived by projecting the 
   combined Python I, II, and III data four-dimensional posterior 
   distributions. The 2 $\sigma$ DMR (marginalized and projected) confidence 
   limits in panels $g)$ and $h)$ are a composite of those from the 
   two extreme DMR data sets (see caption of Figure 4). When the  
   four-dimensional posterior distributions are normalized such
   that $L(Q_{\rm rms-PS}\ =\ 0\ \mu{\rm K})\ =\ 1$, the peak values of
   the one-dimensional distributions shown in panels $a)-h)$ are 
   $2\times 10^{106}$, $3\times 10^{106}$, $6\times 10^{107}$, 
   $7\times 10^{107}$, $1\times 10^{105}$, $2\times 10^{105}$, 
   $9\times 10^{104}$, and $6\times 10^{104}$, respectively.}

\clearpage
\pagestyle{empty}

\begin{center}
  \leavevmode
  \epsfysize=8.0truein
  \epsfbox{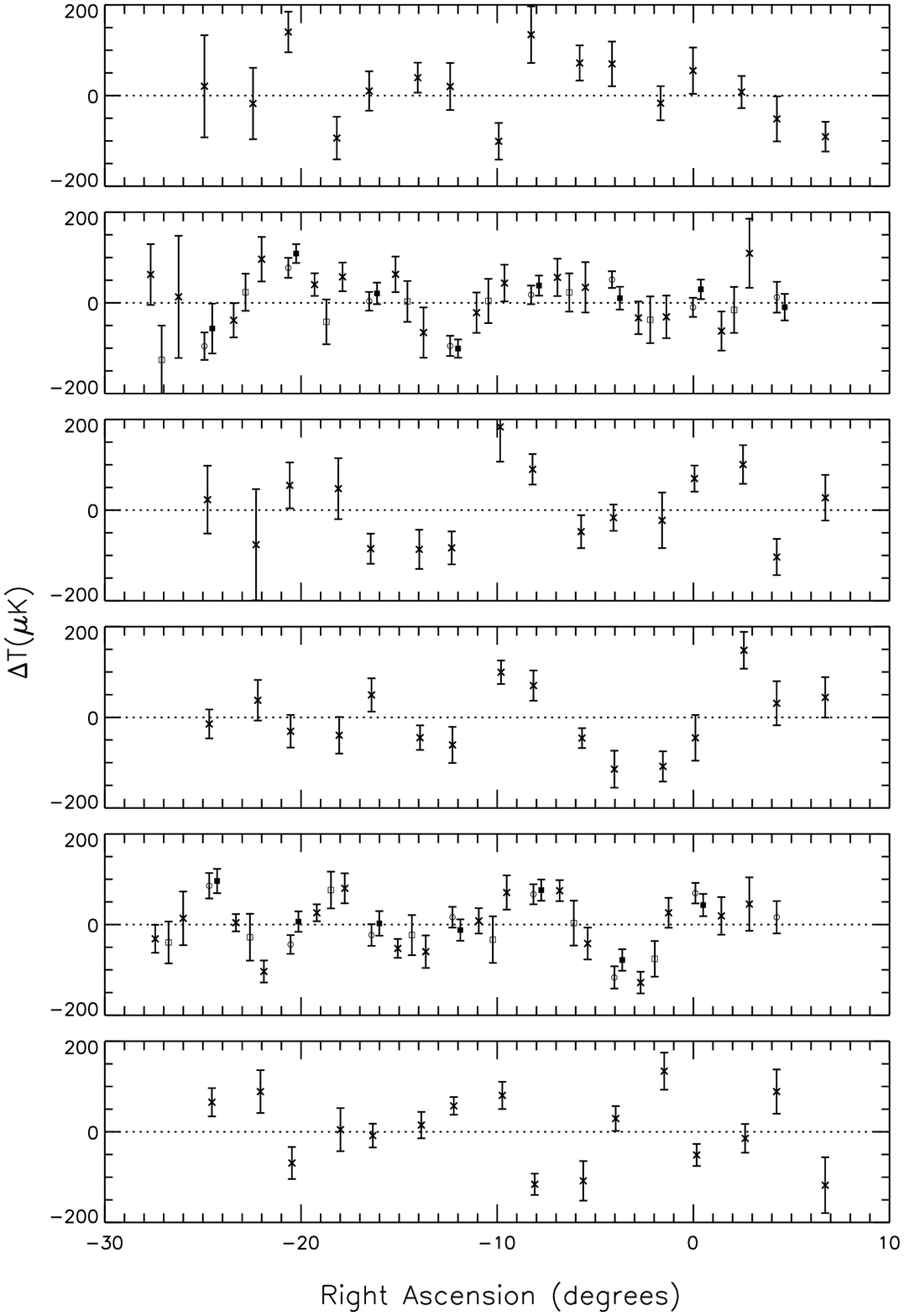}
\end{center}  
\vfill
Figure 1$a)$

\clearpage
\begin{center}
  \leavevmode
  \epsfysize=8.0truein
  \epsfbox{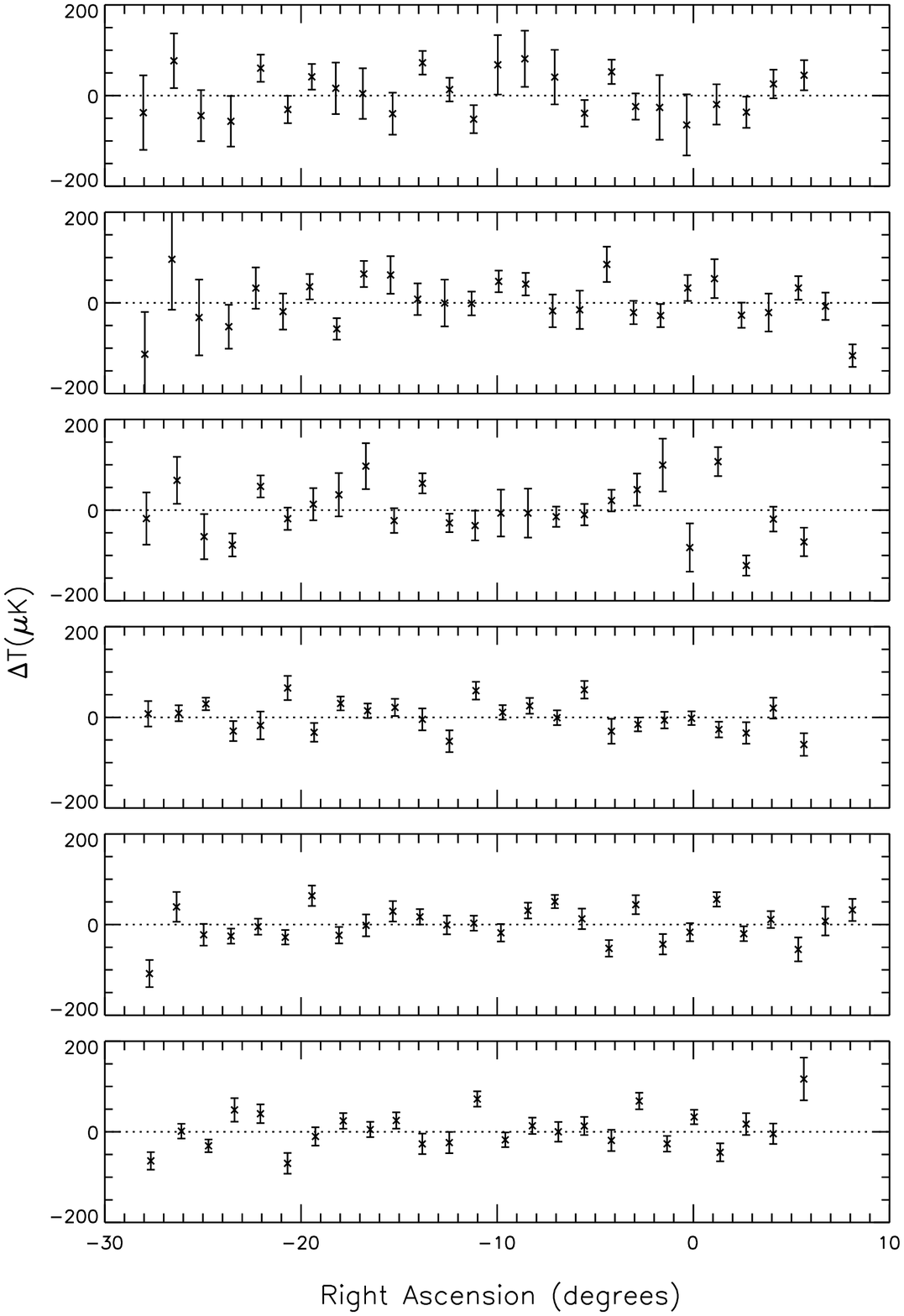}
\end{center}  
\vfill
Figure 1$b)$

\clearpage
\begin{center}
  \leavevmode
  \epsfysize=8.0truein
  \epsfbox{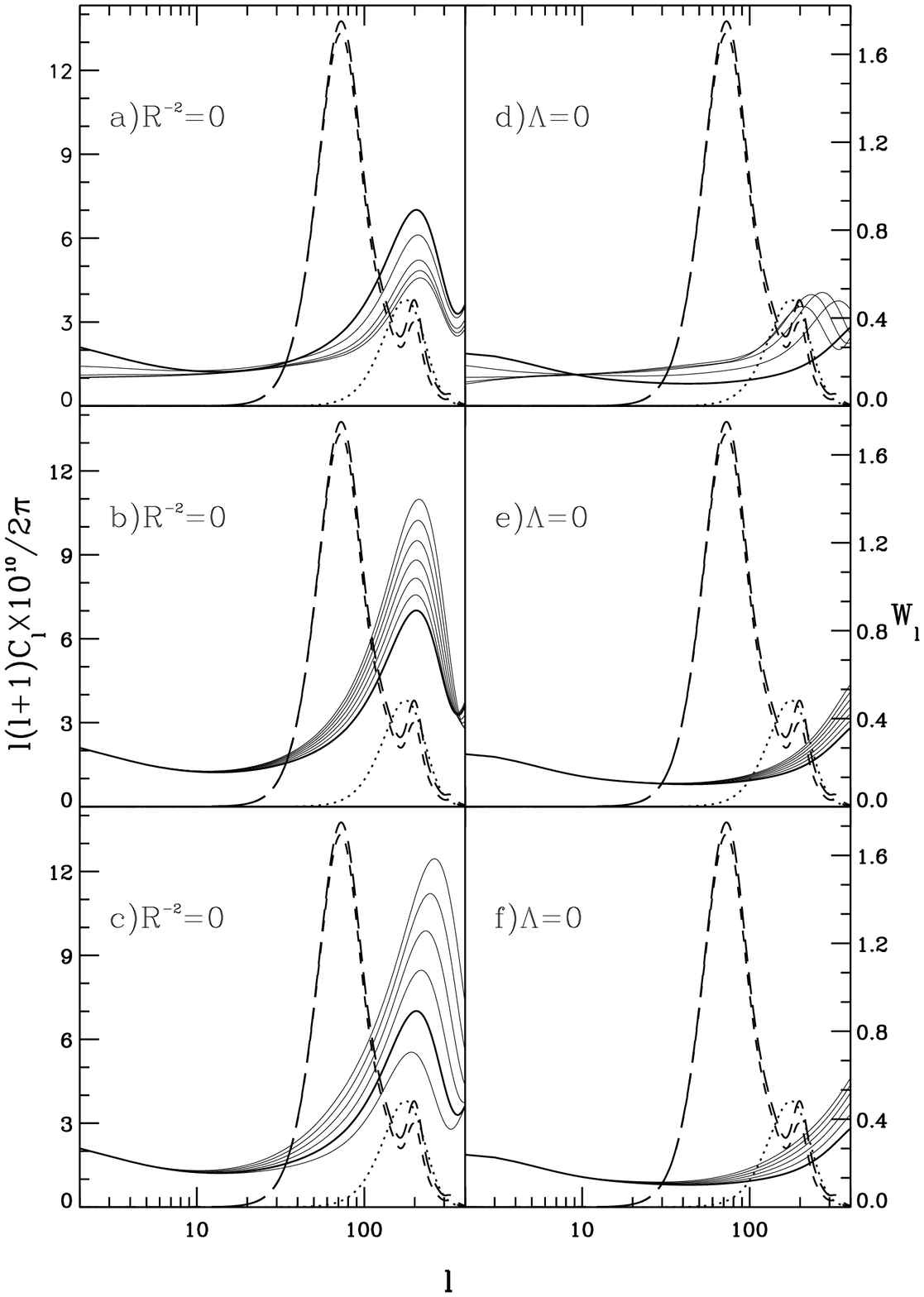}
\end{center}  
\vfill
Figure 2

\clearpage
\begin{center}
  \leavevmode
  \epsfysize=8.0truein
  \epsfbox{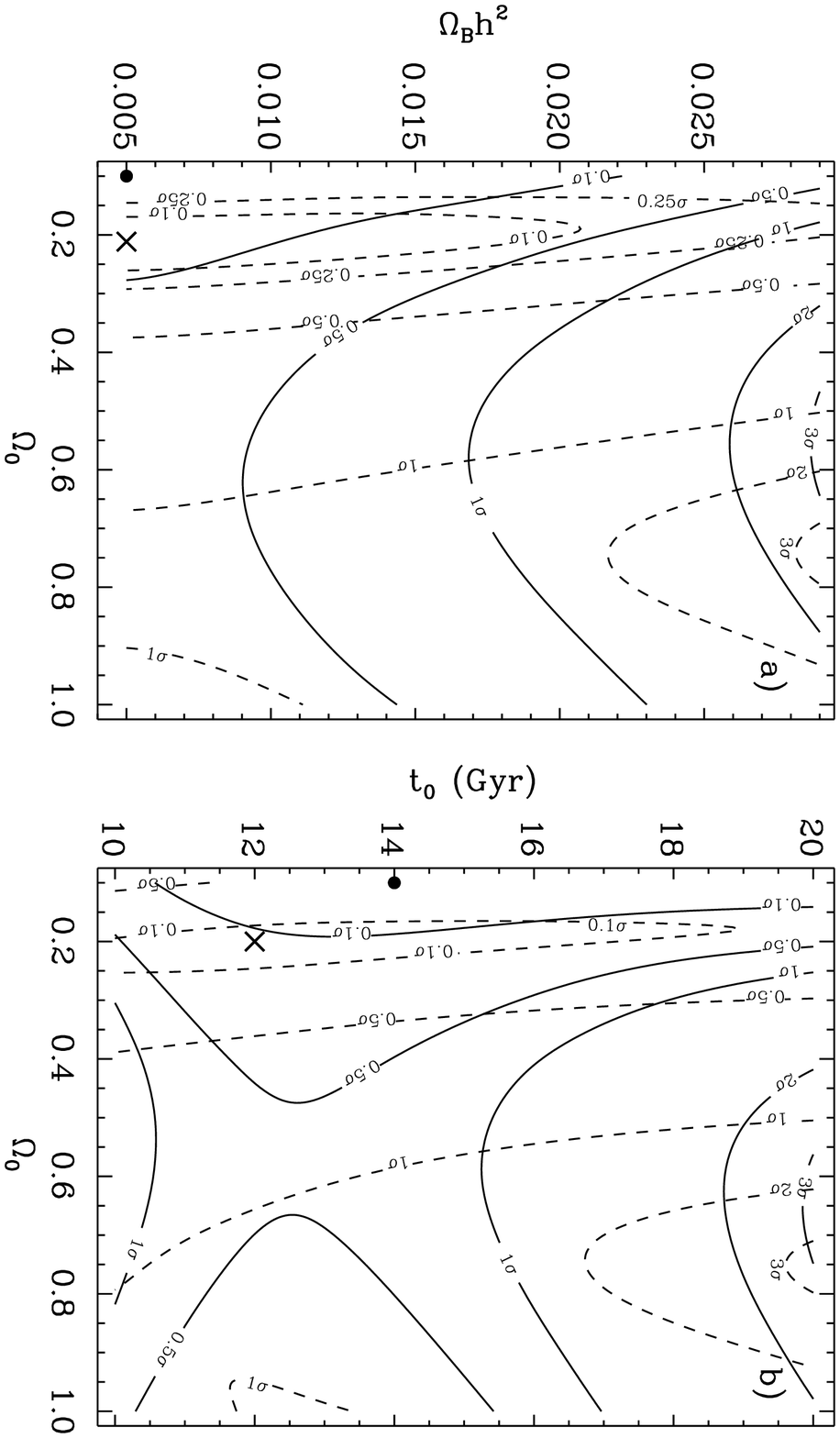}
\end{center}  
\vfill
Figure 3

\clearpage
\begin{center}
  \leavevmode
  \epsfysize=8.0truein
  \epsfbox{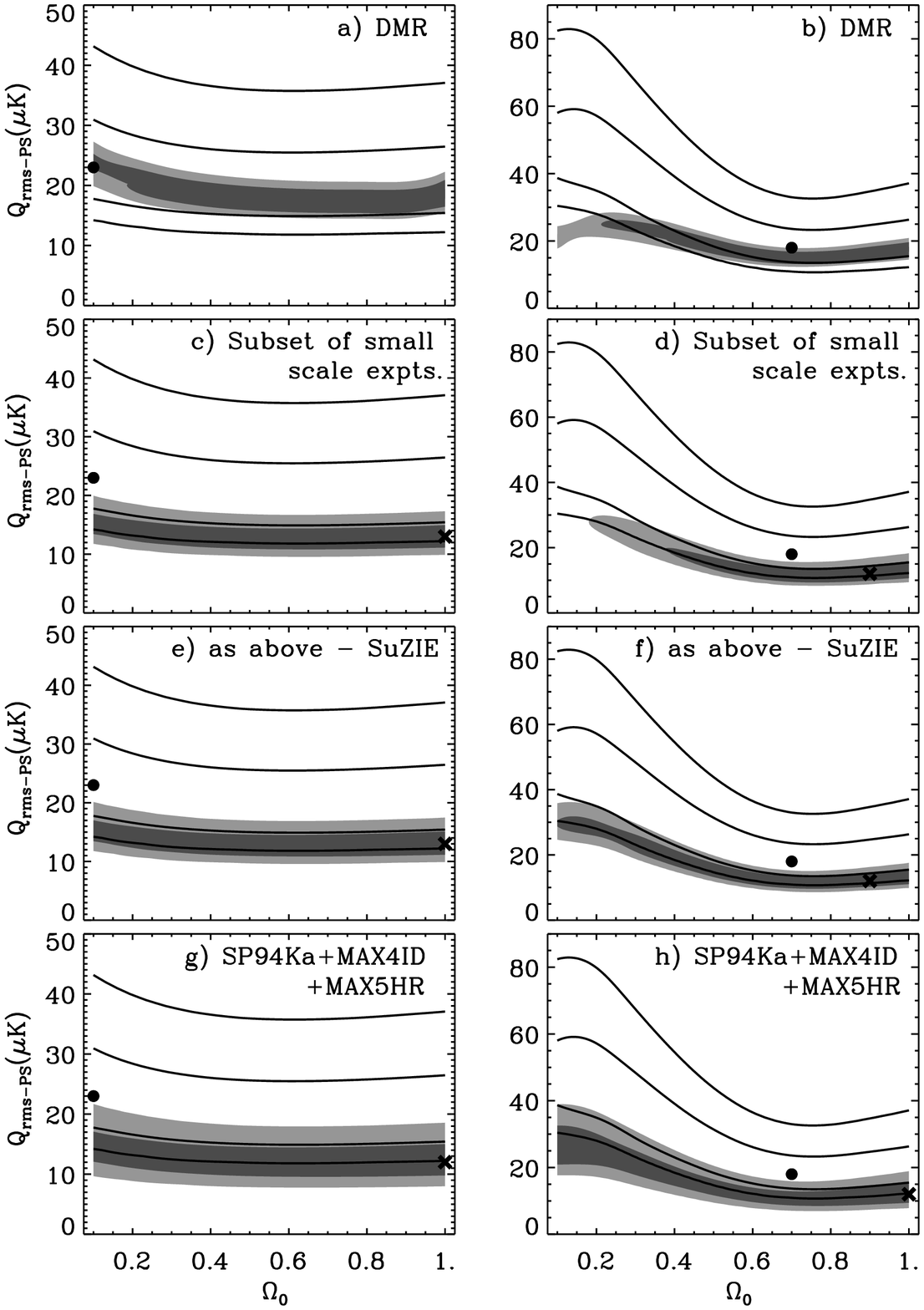}
\end{center}  
\vfill
Figure 4

\clearpage
\begin{center}
  \leavevmode
  \epsfysize=8.0truein
  \epsfbox{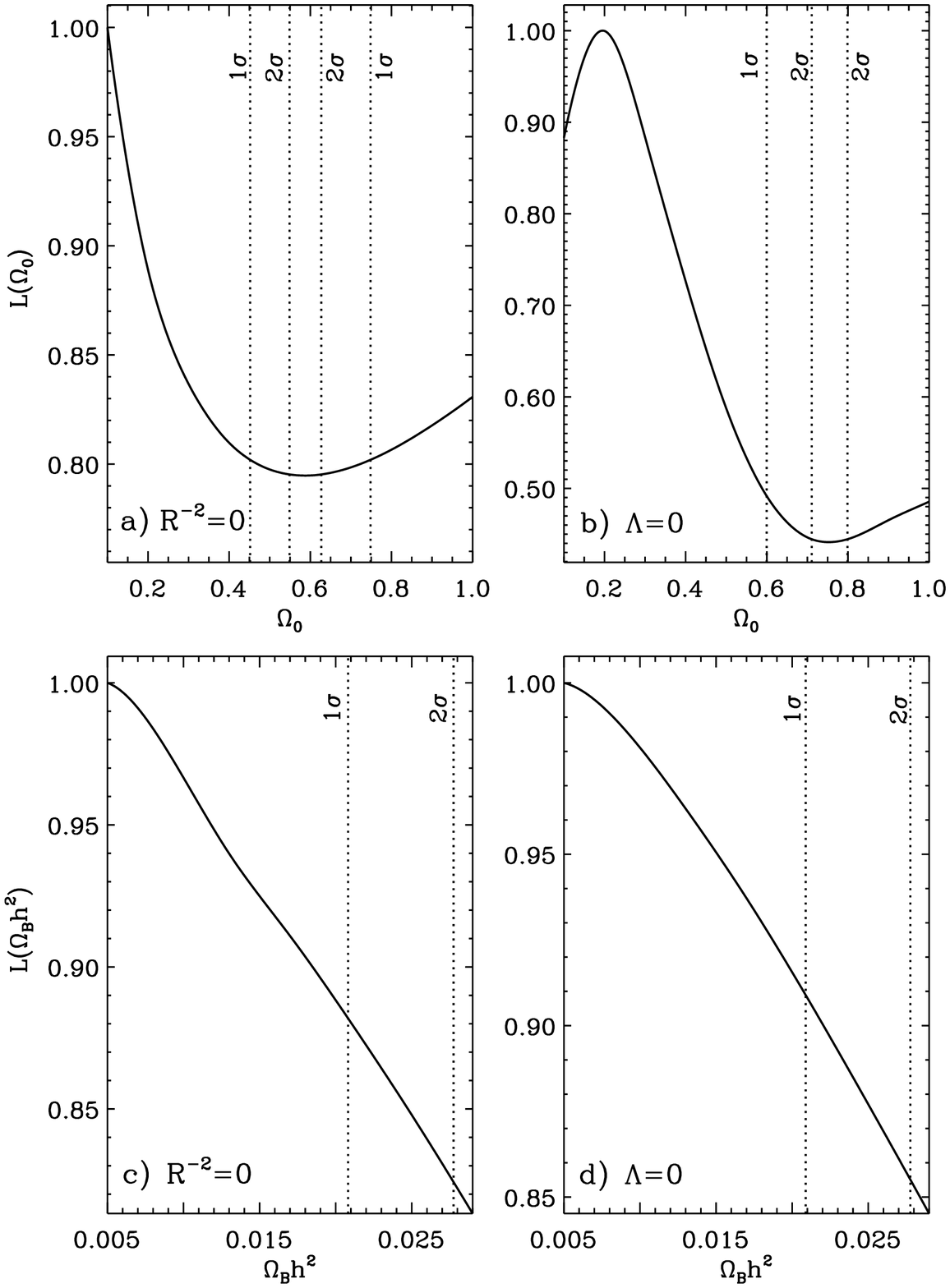}
\end{center}  
\vfill
Figure 5.1

\clearpage
\begin{center}
  \leavevmode
  \epsfysize=8.0truein
  \epsfbox{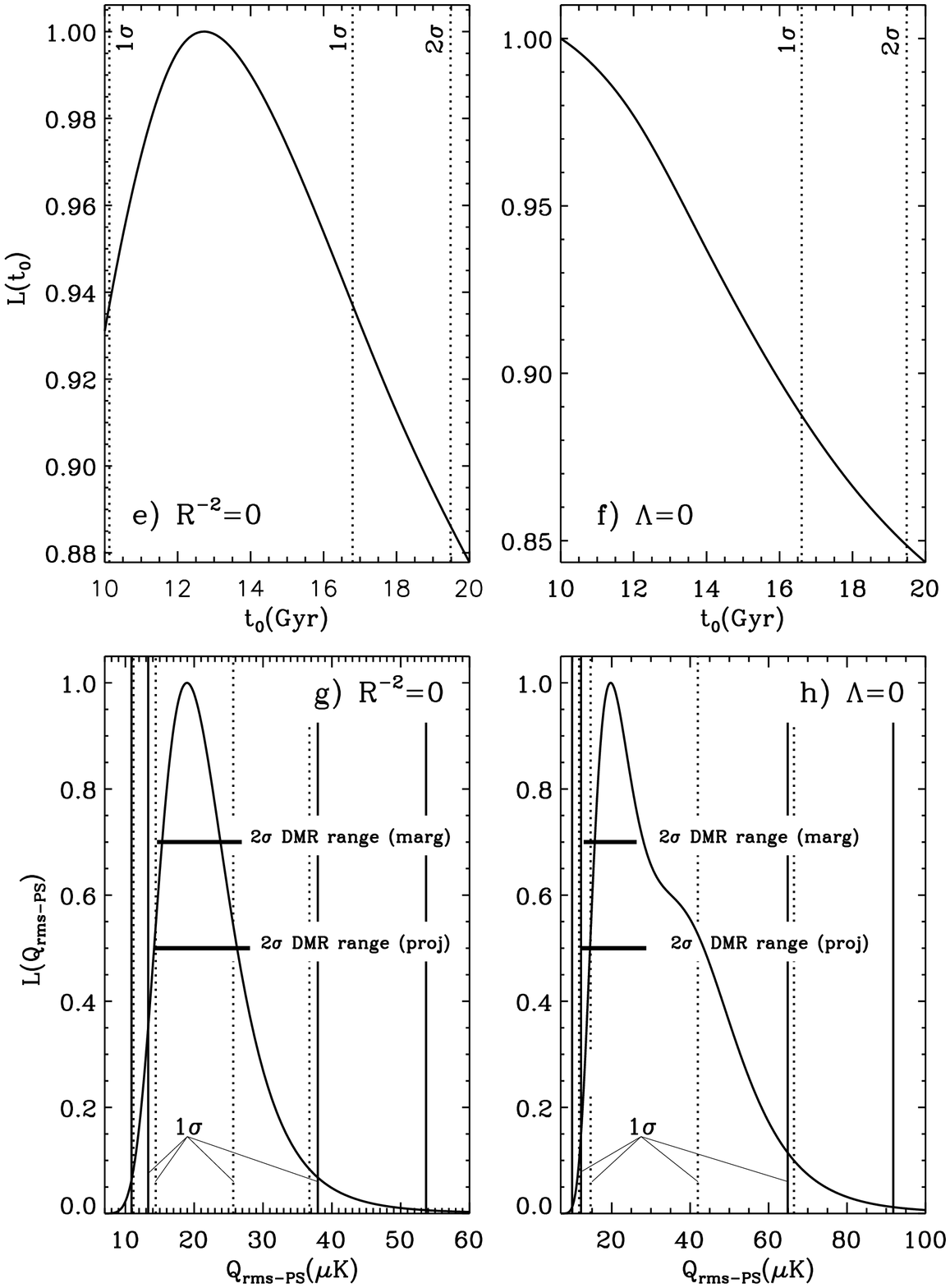}
\end{center}  
\vfill
Figure 5.2

\end{document}